\documentclass[12pt,letterpaper]{article}
\pdfoutput=1
\usepackage{jheppub}
\usepackage{amsfonts, amsthm}
\usepackage[english]{babel}
\usepackage[utf8]{inputenc}
\usepackage{slashed}
\usepackage{mathrsfs}
\usepackage{amssymb}
\usepackage{color}
\hypersetup{unicode}
\usepackage{natbib}

\newcommand{\eq}{\begin{equation}}
\newcommand{\feq}{\end{equation}}
\newcommand{\be}{\begin{equation}}
\newcommand{\ee}{\end{equation}}
\newcommand{\eqn}{\begin{eqnarray}}
\newcommand{\feqn}{\end{eqnarray}}

\newcommand{\mrm}[1]{\mbox{$\mathrm{#1}$}}

\newcommand\secref[1]{Sec.\,\ref{#1}}

\renewcommand{\(}{\left(}
\renewcommand{\)}{\right)}

\title{Nothing really matters}

\author[a]{Giuseppe Dibitetto}
\author[a]{Nicol\`o Petri}
\author[b]{Marjorie Schillo}

\affiliation[a]{Department of Physics, University of Oviedo, Avda. Federico Garcia Lorca s/n, 33007 Oviedo,
Spain.}
\affiliation[b]{Institutionen f\"or fysik och astronomi, University of Uppsala,  Box 803, SE-751 08 Uppsala, Sweden.}

\emailAdd{dibitettogiuseppe@uniovi.es}
\emailAdd{petrinicolo@uniovi.es}
\emailAdd{marjorie.schillo@physics.uu.se}

\abstract{We study non-perturbative instabilities of AdS spacetime in General Relativity with a cosmological constant in arbitrary dimensions. In this simple setup we explicitly construct a class of gravitational instantons generalizing Witten's bubble of nothing. We calculate the corresponding Euclidean action and show that its change is finite. The expansion of these bubbles is described by a lower-dimensional de Sitter geometry within a non-compact foliation of the background spacetime. 
Moreover we discuss the existence of covariantly constant spinors as a possible topological obstruction for such decays to occur. This mechanism is further connected to the stability of supersymmetric vacua in string theory.}

\begin{document}
\maketitle
\flushbottom

\section{Introduction}

The possible decay of metastable spacetimes through gravitational instantons has a rich history and far-reaching applications.  Writ large, the possible instanton decay channels can be grouped as Coleman-de Luccia tunnelling of a scalar field \cite{Coleman:1980aw}, the nucleation of charged membranes via the Brown-Teitelboim mechanism \cite{Brown:1988kg}, and the decay of spacetime itself through Witten's bubble of nothing \cite{Witten:1981gj}.  These processes are of vital import to questions regarding eternal inflation, the ability to populate different corners of the string landscape,  the implications for AdS/CFT in metastable spacetimes,  as well as new cosmological model building possibilities involving braneworld scenarios.

Within the context of string theory in particular, supersymmetry has been argued to be a sufficient condition for non-perturbative stability thanks to the construction outlined in \cite{Witten:1982df}. However, more realistic and phenomenologically appealing string theory constructions must account for a dynamical supersymmetry breaking mechanism. Historically, the first steps in this direction were achieved in a field theory context in \cite{Witten:1998zw} by means of the introduction of antiperiodic boundary conditions for fermions. Subsequently, in \cite{Kachru:2002gs}, a spontaneous supersymmetry breaking mechanism was given a fundamental description using anti-branes. Finally, in \cite{Kachru:2003aw}, the aforementioned construction was coupled to gravity in order to obtain a de Sitter vacuum with broken supersymmetry. In parallel, \cite{Argurio:2006ny} also made use of anti-branes as supersymmetry breaking sources to formulate a non-supersymmetric example of the AdS/CFT correspondence.

Despite these seminal developments in the understanding of non-supersymmetric constructions, their reliability as solutions in a complete theory of quantum gravity has  recently been questioned. The set of  prerequisites  for a low energy effective description to have a consistent completion in quantum gravity  is often referred to as the ``swampland conjectures". The original idea behind this philosophy was presented in \cite{ArkaniHamed:2006dz}, where the role of gravity as the weakest force is investigated. In more recent years a revival of this approach has resulted in a number of (conjectured) consistency conditions that effective theories should satisfy. We address the reader to, e.g. \cite{Brennan:2017rbf, Palti:2019pca} for a nice review of the topic. 

In particular, it was argued in \cite{Ooguri:2016pdq,Freivogel:2016qwc} that all non-supersymmetric, perturbatively stable AdS vacua should suffer from non-perturbative instabilities associated with spontaneous nucleation  of non-extremal charged membranes. The authors of \cite{Ooguri:2016pdq} further speculate on the consequences of these instabilities for the putative  dual field theory of such an AdS vacuum. Despite a tiny probability of a single nucleation event, since the volume of AdS space is infinite, its boundary would decay immediately. This suggests that the idea of a non-supersymmetric holographic correspondence might be problematic.  
However, the study of non-supersymmetric holography within a controlled effective description remains an active topic of reasearch. Early attempts in this direction are discussed e.g. in \cite{Horowitz:1998ha, Balasubramanian:2002am}.

By adopting the approach of the swampland conjectures,  the existence of metastable de Sitter vacua within a consistent theory of quantum gravity has been called into question. Specifically, in \cite{Obied:2018sgi}, it was conjectured that the slope of effective moduli potentials arising from string compactifications can never be small whenever the sign of the corresponding vacuum energy is positive. Additionally, in \cite{Dasgupta:2018rtp,Dasgupta:2019vjn}, no-go theorems were proposed in which controlled dS background geometries are ruled out whenever one restricts to time-independent energy-momentum tensors as sources. If these conjectures are correct, they may be reconciled with current observations  by realizing dark energy in the form of a quintessence supported by time-dependent scalar fields (see e.g. \cite{Blaback:2013fca, Heckman:2019dsj}), or by effectively describing it as a braneworld cosmology which is intrinsically evolving in time.

It is  worth mentioning that,  in the case of non-supersymmetric AdS vacua and holography, as well as in the case of de Sitter  vacua, the issue of their consistency is  far from settled.   Therefore, it is  important to attempt to construct counterexamples which can serve as tests for  conjectured UV consistency requirements. To this end, constructions like those of \cite{Cordova:2018eba} and \cite{Cordova:2018dbb, Cordova:2019cvf} may be important to the above discussion. As a caveat, note that these examples all happen to rely on the existence of a strongly curved region in internal space close to the location of the sources, where the supergravity description may not be reliable.

The main focus of this paper is the study of de Sitter slicings of higher dimensional spacetimes, with or without a cosmological constant, as classical gravitational solutions. From a purely technical viewpoint, constructing this type of backgrounds is  a complicated task due to the necessity of solving second order coupled differential equations without the help of supersymmetry. The toolkit used here consists of the Hamilton-Jacobi formulation, which allows us to recast the original problem into a set of first order conditions even when supersymmetry is broken. The technical details required for the construction are extensively spelled out in Appendices \ref{HJappendix} and \ref{bubbleAdS}.
From a physical perspective, these solutions will provide a description of the semiclassical decay of vacua in arbitrary dimensions via the nucleation of bubbles of nothing. Additionally, these bubble geometries may serve as effective realizations of time-dependent cosmologies through a non-compact foliation of higher dimensional spacetime featuring lower dimensional de Sitter slices.

In particular, we present a family of instantons, generalizing the bubble of nothing, which lead to the decay of $(D+1)$-dimensional vacuum solutions of general relativity with (non-)zero values for the cosmological constant.  As we will review in detail in \secref{bubblenothingwitten}, the bubble of nothing decay results in a compact dimension of the false vacuum shrinking to zero size on the surface of a bubble which subsequently expands.  Therefore the interior spacetime degenerates, and as the bubble expands it destroys the original vacuum.  Bubbles of nothing present a relatively easy way of generating time dependent solutions and have a rich phenomenology that can be extended to string vacua as explored in detail in \cite{Aharony:2002cx}.

It remains an open problem to understand the interpretation of  the decay of asymptotically AdS space in holography (see e.g. \cite{Harlow:2010az, Barbon:2010gn}.)  It is unknown how to interpret the decay in the dual field theory, or even if the decay invalidates the holographic correspondence in unstable spacetimes \cite{Ooguri:2016pdq}.  Significant progress was made in the interpretation of bubble of nothing decays of AdS \cite{Horowitz:2007pr}, and it remains a topic of future work to extend this understanding to other stringy decays such as \cite{Ooguri:2017njy}.  The solutions we present here have the virtue of being extremely simple, and therefore may provide a useful playground for understanding such decay processes.  Furthermore the existence of this new family of decays will be instructive in the ongoing work to extend these solutions to fully stringy decays \cite{Danielsson:2017max, Apruzzi:2019ecr}.

Even though our main focus is the decay of AdS, one can also consider the case of positive cosmological constant. If there exist metastable de Sitter vacua with slow decay rates, these inflating vacua will produce new regions of de Sitter space faster than they can decay, realizing eternal inflation.  In an eternally inflating spacetime any probability to tunnel between different vacua will be realized eventually and the landscape of all connected vacua can be probabilistically explored.  This ability to populate different corners of a landscape is a crucial component of the anthropic solution of the cosmological constant problem \cite{Weinberg:1988cp}.  In this context it is important to understand the relative rates of transition between different vacua.  As their name suggests, bubbles of nothing represent a dead end in the exploration of a landscape and therefore have important consequences for the dynamics of eternal inflation.  Furthermore, it has been suggested that decays via a bubble of nothing could represent the dominant channel in certain flux vacua \cite{Brown:2010mf}. An understanding of possible decay channels and rates is both crucial in order to accept the anthropic solution of the cosmological constant problem, and will be of use in any attempt to place a measure on eternal inflation.

Finally, in the context of  braneworld scenarios, these new decay channels present the possibility of new constructions of a de Sitter phase for low energy observers confined to the bubble wall \cite{Ghosh:2018fbx, Banerjee:2018qey, Banerjee:2019fzz, Amariti:2019vfv}.  Particularly due to their simplicity, the instanton geometries presented here represent useful toy-models to extend the study of ``braneworld'' observers living on ``nothing'' \cite{Yang:2009wz}. This philosophy is in line with a novel interpretation of stringy dS vacua constructed by using infinite throats \cite{Randall:2019ent}.

\section{Instability of the Kaluza-Klein vacuum}
\label{bubblenothingwitten}

In lieu of a full dynamical treatment of the possible ground states of a gravitational system of a theory of quantum gravity, we use the semi-classical approximation to assess possible quantum instabilities of the vacuum. In this context one looks at the decay of a spacetime background as a quantum tunnelling process described by a gravitational instanton. More intuitively, the instanton provides the ``classical picture" of a quantum process by which a bubble spontaneously appears and expands in spacetime.  This process is characterized by a finite probability for nucleation and furthermore describes the subsequent time-dependent dynamics which describe the conversion of the initial ``false vacuum'' into a new state inside the bubble.

In this section we review the original results of Witten's bubble of nothing, \cite{Witten:1981gj}, where the five-dimensional Kaluza-Klein (KK) vacuum, $\mathbb{R}^{1,3}\times S^1$, is shown to have a non-perturbative instability.\footnote{This is done purely for pedagogical purposes since the main results of this paper are direct generalizations of  \cite{Witten:1981gj}. Readers already familiar with  \cite{Witten:1981gj} are invited to skip to the next section.}  Although this spacetime is a vacuum solution to Einstein's equations with zero energy, it has different asymptotic topology from the five-dimensional Minkowski vacuum, $\mathbb{R}^{1,4}$.  Therefore, a comparison of energies at infinity is meaningless between these two spacetimes, and the positive energy theorem \cite{Schon:1979rg,Schon:1981vd,Witten:1981mf}, which ensures the full stability of $\mathbb{R}^{1,4}$ does not apply to $\mathbb{R}^{1,3}\times S^1$.

It follows that instead of comparing asymptotically distinct spacetimes, one should simply search for gravitational instantons describing the decay of the KK vacuum. Such an instanton should be a real, smooth solutions of the Euclidean equations of motions approaching the false vacuum at asymptotic infinity.  In order for such a solution to contribute to the semiclassical path integral it should additionally have a finite Euclidean action $S_E|_{\text{inst}}$. Then, the quantum tunnelling process is possible with the decay rate per volume of spacetime given by \cite{Coleman:1977py}
\begin{equation}
\begin{split}
 \Gamma/V &\propto e^{-\Delta S_E}\,, \\
 \Delta S_E&=S_E|_{\text{inst}}-S_E|_{\text{false}} \,
 \end{split}
 \label{bouncedef}
\end{equation}
where $\Delta S_E$ is often called the ``bounce''.

The key idea of \cite{Witten:1981gj} is to use a double-analytic continuation of non-extremal black holes in five dimensions to construct instantons in the KK vacuum. In particular, beginning with the five-dimensional Schwarzschild black hole,
\begin{equation}
 ds^2=-\left(1-\(\frac{R}{\rho}\)^2\right)dt^2+\frac{d\rho^2}{\left(1-\(\frac{R}{\rho}\)^2\right)}+\rho^2ds^2_{S^3}\,,
\end{equation}
 one can find a non-singular Euclidean solution by Wick-rotating the time coordinate  $t\rightarrow i \phi$, 
\begin{equation}
 ds^2_{E}=\left(1-\(\frac{R}{\rho}\)^2\right)d\phi^2+\frac{d\rho^2}{\left(1-\(\frac{R}{\rho}\)^2\right)}+\rho^2ds^2_{S^3}\,,
 \label{eucSchwarzschild}
\end{equation}
However, this solution is well-defined and smooth only for a radial coordinate $\rho\in (R, +\infty)$ and when the coordinate $\phi$ is periodic with $\phi \sim \phi + 2\pi R$. Clearly, in the asymptotic limit, $\rho\rightarrow +\infty$, the solution \eqref{eucSchwarzschild} approaches the Euclidean KK vacuum written in spherical coordinates, \emph{i.e.}
\begin{equation}
ds^2_{\text{EKK}}=d\phi^2+d\rho^2+\rho^2ds^2_{S^3}\,,
\label{euclideanKK}
\end{equation}
where the KK radius, $\ell_{\text{KK}}$, is identified with the period of the $\phi$ coordinate of the Euclidean black hole, $R=\ell_{\text{KK}}$. 

The last requirement for this Euclidean solution to describe a bubble nucleation is that it can be continued back to a real metric with Lorentzian signature along a time-symmetric plane identified with $\tau=0$.  Looking at the asymptotic form of the Euclidean metric \eqref{euclideanKK}, it is clear that the analytic continuation has to be performed inside the $S^3$ if we want to keep intact the asymptotic structure of the KK vacuum.  Moreover, parametrizing the $S^3$ in the usual way, $ds^2_{S^3}=d\theta^2+\sin^2\theta\,ds_{S^2}^2$, the equator of the $S^3$ provides a symmetric $\tau=0$ plane via the analytic continuation $\theta\rightarrow i\tau+\frac{\pi}{2}$. In this way $ds^2_{S^3}\rightarrow ds^2_{{\scriptsize \mrm{dS}_3}}=-d\tau^2+\cosh^2\tau \,ds^2_{S^2}$ and the instanton metric \eqref{eucSchwarzschild} becomes
\begin{equation}
 ds^2_{5}=\left(1-\left(\frac{R}{\rho}\right)^2\right)d\phi^2+\frac{d\rho^2}{1-\left(\frac{R}{\rho}\right)^2}+\rho^2ds^2_{{\scriptsize \mrm{dS}_3}}\,.
 \label{bubblenothing5d}
\end{equation}
This background retains the restriction on the radial coordinate $\rho\in(R,+\infty)$, and is non-singular and geodesically complete.

In the asymptotic limit, $\rho\rightarrow +\infty$, we recover the KK vacuum, however in the interior, the presence of the Schwarzschild warp factors and the range of $\rho\in(R, +\infty)$ indicate that the $(\rho,\tau)$-plane parametrizes the exterior region of a hyperbola of radius $R$.  This hyperbola is exactly a de Sitter manifold $\mrm{dS}_3$ which describes a bubble expanding under constant proper acceleration.  At the surface $\rho=R$ the KK direction, $\phi$ pinches off to zero size, so that the interior of the bubble is dubbed ``nothing.''

Therefore, the solution \eqref{bubblenothing5d} constitutes an instanton describing a non-perturbative instability of the KK vacuum which leads to a bubble of nothing expanding on a de Sitter hyperbola and eventually ``eating'' the entire KK vacuum. The fact that this decay will happen with probability equal to one in an infinite volume of space can be seen by computing a finite  difference in Euclidean actions
\begin{equation}
 \Delta S_E=\frac{\pi R^2}{4\,G_N^{(4)}}\,.
 \label{bnothingpprob5d}
\end{equation}
Since this analysis has been performed in the semi-classical approximation, we can only trust the interpretation of the decay for values of $R$ that are large with respect to the Planck length, $\ell_{\text{Pl}}$.

\section{Bubbles of nothing and vacuum decay}
\label{bubbleKK}

In this section we extend the analysis of \cite{Witten:1981gj} to a larger class of instabilities in  constant curvature vacuum solutions in $(D+1)$-dimensional General Relativity \cite{Godazgar:2009fi}. These vacua differ from the ``standard'' KK vacua, such as the $\mathbb{R}^{1,3}\times S^1$ vacuum of \cite{Witten:1981gj}, in their asymptotic topology. In addition to the presence of a cosmological constant $\Lambda$, the main difference is an asymptotic region described not by one ``celestial sphere'' but rather a product of spheres in the corresponding Euclidean vacua.  The zero-curvature case will describe a Ricci flat, zero-energy, vacuum solution and it will directly extend the 5d case studied in \cite{Witten:1981gj}\footnote{Note that this is not possible for $D\le4$ with $\Lambda=0$ due to no-hair theorems, however these do not apply for $\Lambda \ne 0$ or in higher dimensions as can be seen by a zoology of solutions \emph{e.g.}\,collected in \cite{Emparan:2008eg}.}.
In the case of non-zero curvature we will  find asymptotically locally  $\mathrm{AdS}$($\mathrm{dS}$) bubble geometries which asymptote to  metastable vacua of the form
\eq
\begin{split}
ds^2_{D+1}=&\(1-k \rho^2\)d\phi^2 + \({D-2\over d-1}\) {d\rho^2 \over \(1-k \rho^2\)} + \rho^2\(L^2_{{\rm dS}_d}ds^2_{{\rm dS}_d} + L^2_{S^{D-d-1}}ds^2_{S^{D-d-1}}\)\,, \\
k=& {2(D-2) \over D(D-1)(d-1)}\Lambda\,,
\end{split}
\label{falseDplus1}
\feq
where $\phi$ is the compact direction describing the KK circle with period\footnote{Note that while this $S^1$ is not technically compact in the case with non-zero curvature, we will still refer to it as a KK circle.} $2\pi \ell_{\text{KK}}$, and $ds^2_{{\rm dS}_d}\,(ds^2_{S^{D-d-1}})$ is the line element for a $d$-dimensional de Sitter (($D-d-1$)-dimensional sphere) with unit radius.  The cosmological constant, $\Lambda$, measures the constant curvature of the spacetime and is related to the Ricci scalar, ${\cal R}$, for a given curvature radius, $\ell$, via
\be
|\Lambda| = {D(D-1) \over 2 \ell^2}, \qquad {\cal R}={2(D+1) \over D-1} \Lambda\,.
\ee
These solutions represent non-singular and geodesically complete vacuum solutions ($R_{\mu\nu}={2 \over D-1}\Lambda g_{\mu\nu}$) when the following algebraic conditions
\be
L_{{\rm dS}_d}=1, \qquad  L_{S^{D-d-1}}=L_{{\rm dS}_d}\sqrt{{D-d-2 \over d-1}}\,.
\label{radiiconstraint}
\ee
are satisfied. 
Note that we need $d\ge2$ and $D\neq d+2$ for the full $(D+1)$-dimensional solution to exist.

We point out that the case $D=d+1$ is peculiar. This correspond in fact to the case in which the foliation with $S^{D-d-1}$ is not present. In this case the vacuum \eqref{falseDplus1} reproduce a AdS topology when $\Lambda$ is negative. When the additional $S^{D-d-1}$ is present the geometry is only locally $\mathrm{AdS}_{D+1}$. Moreover, as we will see, the 5d case studied in \cite{Witten:1981gj} can be recovered when the second sphere is absent.

Continuing to a Euclidean metric, in analogy to \secref{bubblenothingwitten}, we notice that the same asymptotic behavior can be achieved in the $\rho \to \infty$ limit of  a Euclidean black hole whose horizon is the product of two spheres,
\begin{equation}
\begin{split}
 ds^2_{E,D+1}=& f(\rho)d\phi^2+\left(\frac{D-2}{d-1}  \right)\frac{d\rho^2}{f(\rho)} +\rho^2\(ds^2_{S^d} + \({D-d-2 \over d-1}\) ds^2_{S^{D-d-1}}\)\,, \\
f(\rho)=&1 -k \rho^2 - \({R\over \rho}\)^{D-2}\,.
\end{split}
 \label{bubblenothingeuclid}
\end{equation}
In the Lorentzian black hole solution, there is a coordinate singularity at $\rho=\rho_0$, where $f(\rho_0)=0$. In order to ensure a smooth solution at $\rho = \rho_0$ we examine the  $\rho \rightarrow \rho_0$ limit where the metric \eqref{bubblenothingeuclid} takes the form
\begin{equation}
\begin{split}
 ds^2_{E,D+1}\approx&\left(\frac{4(D-2)}{(d-1)f^\prime (\rho_0)} \right)\left(d \zeta^2 +\frac{(d-1)f^\prime(\rho_0)^2}{4(D-2)}\zeta^2\,d\phi^2 \right)\\
 &+ \rho_0^2\(ds^2_{S^d} + \({D-d-2 \over d-1}\) ds^2_{S^{D-d-1}}\)\,,
 \end{split}
 \label{surfacebubbleAdS}
\end{equation}
where $\zeta=(\rho-\rho_0)^{1/2}$.  It follows that there will be no conical defect in the $(\zeta,\phi)$ plane if we impose the condition
\begin{equation}
\ell_{\text{KK}}^2 = {4(D-2) \over f'(\rho_0)^2(d-1)}\,.
\label{lkktorho0cond}
\end{equation}
For the case $\Lambda \le 0$ this leaves the Euclidean solution \eqref{bubblenothingeuclid} well-defined and non-singular for $\rho \in (\rho_0, +\infty)$.  In the case of $\Lambda >0$, due to a zero in the function $f'(\rho)$ and the use of the static patch of de Sitter, the Euclidean geometry does not have an asymptotic region.  Henceforth, we will specialize to the case $\Lambda \le 0$. 

In direct analogy to the construction of bubble of nothing in \cite{Witten:1981gj}, we  see that continuing $S^d$ to dS$_d$ will produce an instanton geometry describing an expanding bubble whose surface is $\mrm{dS}_d\times S^{D-d-1}$
\begin{equation}
ds^2_{D+1}=f(\rho)d\phi^2+\left(\frac{D-2}{d-1}  \right)\frac{d\rho^2}{f(\rho)} +\rho^2\(ds_{{\rm dS}_d}^2 + \({D-d-2 \over d-1}\) ds^2_{S^{D-d-1}}\)\,.
 \label{bubblenothingKK}
\end{equation}
In Appendix \ref{bubbleAdS} the direct derivation of these solutions is furnished by providing a first-order formulation of these $\mathrm{dS}_d$ backgrounds through the Hamilton-Jacobi formulation sketched in Appendix \ref{HJappendix}.
The asymptotic geometry of these solutions reproduces the vacuum \eqref{falseDplus1} and the geometry is defined for values $\rho \in (\rho_0, + \infty)$.  Thus we have a generalization of the bubble of nothing describing the decay of the vacuum solutions \eqref{falseDplus1}.

Notice that in the zero-curvature case the relation \eqref{lkktorho0cond} simplifies to
\begin{equation}
 \ell_{\text{KK}}^2 = {4 R^2 \over (d-1)(D-2)}\,,\label{lKKzerocurvature}
\end{equation}
where $\rho_0=R$.
The case of \cite{Witten:1981gj} is thus recovered by imposing
\begin{equation}
\begin{split}
 &D=4\,,\qquad d=3\quad\Longrightarrow \quad  R=\ell_{\text{KK}}\,.
 \end{split}
\end{equation}



\subsection{Euclidean action}
\label{euclidactionKK_section}

In order to establish that the bubble geometry \eqref{bubblenothingKK} describes a non-perturbative instability of the vacuum \eqref{falseDplus1} we need to show that the change in Euclidean action is finite. 
It is well known that the the variational problem for gravity is not completely fixed by the requirement of vanishing variations of the gravitational field at the boundary. In particular, in order to reproduce Einstein equations one has to include into the action the Gibbons-Hawking-York (GHY) term that takes in account those boundary contributions that are not fixed by the requirement of vanishing variations at the boundary. In our considerations, both the bubble of nothing in the instanton geometry and the conformal boundary at spatial infinity constitute boundaries where the GHY term must be evaluated.

The Euclidean action is given by
\begin{equation}
 \begin{split}
  &S_E=S_{E, \,\text{bulk}}+S_{E,\, \text{GHY}}\,,\\
  &S_{E,\, \text{bulk}}=\frac{1}{2\kappa_{D+1}^2}\int{d^{D+1} x_E\,\sqrt{g_{D+1}}\,\left ({\cal R}-2\Lambda \right)}\,,\\
  &S_{E, \,\text{GHY}}=\frac{1}{\kappa_{D+1}^2}\int{d^{D} y_E\,\sqrt{h_{D}}\,\theta_{D}}\,,
 \end{split}\label{euclidactionKK}
\end{equation}
where in the GHY boundary term, the coordinates $\{y\}$ parametrize any $D$-dimensional boundary of the background described by the ``bulk action" $S_{\text{bulk}}$, $h_{D}$ is the induced metric and $\theta_{D}$ is the trace of the extrinsic curvature of the boundary. 

We begin by computing $S_{E,\text{bulk}}$ for the metric \eqref{bubblenothingeuclid} obtaining the following expression
\be
\begin{split}
S_{E,\text{bulk}} &= {1\over 2\kappa_{D+1}^2} \int d\phi\, d\rho\, d\Omega_d\, d\Omega_{D-d-1} \sqrt{{D-2\over d-1}}\({D-d-2 \over d-1}\)^{(D-d-1)/2} \rho^{D-1} {4\Lambda\over D-1}    \, \\
&= {\pi \ell_{\text{KK}} \over \kappa_{D+1}^2} {\rm vol}_{S^d} {\rm vol}_{S^{D-d-1}}\sqrt{{D-2\over d-1}}\({D-d-2 \over d-1}\)^{(D-d-1)/2} {4\Lambda\over D-1} \int d\rho \rho^{D-1}  \,,
\end{split}
\label{SEbulk}
\ee 
where $d\Omega_n$ is the measures on an n-spheres $S^n$.
Note that this expression is valid for both the Euclidean false vacuum \eqref{falseDplus1}, and also the instanton geometry \eqref{bubblenothingeuclid}, however the limits of integration will differ.  Particularly, computing  the bulk contribution to the bounce \eqref{bouncedef}, we find
\be
\Delta S_{E,\text{bulk}} = C\(\int_{\rho_0}^\infty d\rho \rho^{D-1} - \int_0^\infty d\rho \rho^{D-1}\) = -C\,{\rho_0^D \over D}   \,,
\label{bouncebulk}
\ee
where $C$ is the prefector to the integral in \eqref{SEbulk}.

Now we compute the contribution from boundary terms.  Both the bubble of nothing and the asymptotic boundaries are surfaces of constant $\rho$. Therefore all contributions of the GHY term to the action can be computed via induced metrics $h_{\mu \nu}=g_{\mu \nu} - \hat n_\mu \hat n_\nu$ and the trace of the extrinsic curvature on surfaces normal to $\hat n_\mu = \hat \rho$. This gives a general boundary term
\be
\begin{split}
S_{E,\text{GHY}} =& {1\over \kappa_{D+1}^2}\int d\phi d\Omega_d\, d\Omega_{D-d-1}\({D-d-2 \over d-1}\)^{(D-d-1)/2} \sqrt{f(\rho)}\rho^{D-1} \\
& \times \sqrt{{d-1\over D-2}} \({ -2(D-1) + D\({R\over \rho}\)^{D-2} + 2D k\rho^2  \over 2\rho \sqrt{f(\rho)} }\) \,,
\end{split}
\ee
where the quantity on the second line is $\theta_D$.  Computing the boundary contributions to the bounce, we note that at the asymptotic boundary, in the $\rho \to \infty$ limit, all dependence on $R$ vanishes and asymptotic contributions to the bounce (which are divergent in the case of non-zero curvature) will cancel between the false vacuum and the instanton solution.   Furthermore, the contribution at the surface of the bubble of nothing should be taken with a reversed sign to take into account the fact that the outward facing normal is $-\hat\rho$.  This gives a bounce contribution
\be
\begin{split}
\Delta S_{E,\text{GHY}} =& - S_{E,\text{GHY}}|_{\rho_0} \,\\
=& -{\pi \ell_{\text{KK}} \over \kappa_{D+1}^2} {\rm vol}_{S^d} {\rm vol}_{S^{D-d-1}}\sqrt{{d-1\over D-2}}\({D-d-2 \over d-1}\)^{(D-d-1)/2}  \\
&\times \rho_0^{D-2} \(-2(D-1) + D\({R\over \rho_0}\)^{D-2} + 2Dk\rho_0^2\)\,.
\end{split}
\label{bouncebndry}
\ee
At this point we consider the cases of zero and non-zero curvature separately.
\paragraph{Zero curvature case: $\Lambda = 0$.}  In this case \eqref{bouncebulk} vanishes, since it is proportional to $\Lambda$, and  $\rho_0 =R$.  Then the bounce is given by
\be
\Delta S_E^{(\Lambda=0)} =   {\pi \ell_{\text{KK}} \over \kappa_{D+1}^2} {\rm vol}_{S^d} {\rm vol}_{S^{D-d-1}}\sqrt{{d-1\over D-2}}\({D-d-2 \over d-1}\)^{(D-d-1)/2} (D-2)R^{D-2} \,.
\ee
If we consider the case $D=4$ and $d=3$, the $(D-d-1)$-sphere disappears, ${\rm vol}_{S^d} = 2\pi^2$ and we reduce to the five-dimensional result \eqref{bnothingpprob5d} of \cite{Witten:1981gj} using the standard relation between the Planck masses under dimensional reduction
\be
{1\over \kappa_D^2} = {1 \over 8 \pi G_D} = {2\pi \ell_{\text{KK}} \over \kappa_{D+1}^2} \, .
\ee 

It is instructive to consider the various length scales which contribute to this decay rate.  Ignoring $D$- and $\pi$-dependent factors and using $\kappa_{D+1}^{2} \propto \ell_{\text{Pl}}^{D-1}$, where $\ell_{\text{Pl}}$ is the $(D+1)$-dimensional Planck length, we can write
\be
\Delta S_E^{(\Lambda=0)} \propto {\ell_{\text{KK}} \over \ell_{\text{Pl}}} \({R \over \ell_{\text{Pl}}}\)^{D-2} \propto \({R \over \ell_{\text{Pl}}}\)^{D-1}\, ,
\ee
where in the second relation we have used $\rho_0=R$ and \eqref{lKKzerocurvature}.  Thus, we have a manifestly positive bounce that will be large whenever the semiclassical approximation is valid, giving rise to an exponentially suppressed decay rate.
\paragraph{Non-zero curvature case: $\Lambda < 0$.} In this case we have contributions both from the bulk and boundary in the bounce; combining \eqref{bouncebulk} and \eqref{bouncebndry} we find the following
\be
\begin{split}
\Delta S_E^{(\Lambda\neq0)} =& {\pi \ell_{\text{KK}} \over \kappa_{D+1}^2} {\rm vol}_{S^d} {\rm vol}_{S^{D-d-1}}\sqrt{{d-1\over D-2}}\({D-d-2 \over d-1}\)^{(D-d-1)/2} \\
&\times \Big[2(D-1)\rho_0^{D-2} - 2(D+1) k \rho_0^D -D R^{D-2}\Big] \\
=&  \,{\pi \ell_{\text{KK}} \over \kappa_{D+1}^2} {\rm vol}_{S^d} {\rm vol}_{S^{D-d-1}}\sqrt{{d-1\over D-2}}\({D-d-2 \over d-1}\)^{(D-d-1)/2} \\
&\times \Big[(D-2)R^{D-2} - 4k\rho_0^D\Big] \,,
\end{split}
\ee
where in passing to the second equality we have rewritten the term in square brackets using the fact that $f(\rho_0)=0$.  We see that the first term reproduces the zero-curvature case and the second term gives a contribution proportional to $\Lambda$.

Looking at the scales involved we find
\be 
\Delta S_{E}^{(\Lambda\neq 0)} \propto {\ell_{\text{KK}} \over \ell_{\text{Pl}}} \( \({R \over \ell_{\text{Pl}}}\)^{D-2} + {4\over D-2} \({\rho_0 \over \ell_{\text{Pl}}}\)^{D-2} \({ \rho_0 \over \ell}\)^2 \) \, .
\ee
Hence, for the case $\Lambda <0$, the decay rate is always exponentially suppressed when the semiclassical approximation is valid. 

\subsection{Properties of the decay}
\label{discussion}

First, consider the zero-curvature case by recalling some energy considerations of \cite{Witten:1981gj}. 
Since any decay process has to preserve energy, it follows that the 5d spacetime \eqref{bubblenothing5d} must have zero energy and this means that a positive energy theorem for the $\mathbb{R}^{1,3}\times S^1$ vacuum does not exist: there exists a solution with zero energy asymptotically approaching $\mathbb{R}^{1,3}\times S^1$, and this is exactly what allows the vacuum to decay. This fact can be seen by looking at the surface integral defining the energy. In particular, taking the zero-time surface at $\tau=0$, the asymptotics of \eqref{bubblenothing5d} gives contributions to the geometry of $\mathbb{R}^{1,3}\times S^1$ of the order $\rho^{-2}$ and it can be shown that, for a four-dimensional observer, only terms of the order $\rho^{-1}$ give positive contributions to the energy of the vacuum. This argument can be extended to our $(D+1)$-dimensional backgrounds \eqref{bubblenothingKK}.
For a $(D>4)$-dimensional observer, the relevant contributions to the energy of the false vacuum \eqref{falseDplus1} have to be of the order $\rho^{-(D-3)}$.  Therefore, energy is conserved during the decay since \ corrections coming from the bubble geometry \eqref{bubblenothingKK} are of the order $\rho^{-(D-2)}$.

The above discussion is  related to the intrinsic KK structure of the vacuum \eqref{falseDplus1}. The radius of the compact dimension $\ell_{\text{KK}}$, which provides an infrared cut-off for the $(D+1)$-dimensional theory, defines a UV cut-off for an  effective lower-dimensional theory. In particular, the physics experienced by a $D$-dimensional observer will be governed by gravity and a running scalar\footnote{See Appendix \ref{bubbleAdS} for the detailed analysis.}. In the asymptotic regime of the bubble geometry, the $D$-dimensional background is given by a Minkowski spacetime without the portion corresponding to the interior of the hyperbola with radius $R$. Moving into the ``bulk" and approaching the boundary, this observer experiences a singular geometry. This singular behavior is consistently resolved by taking the point of view of a $(D+1)$-dimensional observer who sees a KK geometry asymptotically and a smooth geometry in the ``bulk". The corrections to the energy of the $(D+1)$-vacuum \eqref{falseDplus1} coming from \eqref{bubblenothingKK} will then give positive contribution to the ``incomplete" effective $D$-dimensional theory, and meanwhile, they will leave  the energy of the vacuum of the higher-dimensional theory intact.

If we include in this discussion  the bubble geometries with constant curvature, we point out that the key property characterizing these decays is the change of topology \cite{Witten:1981gj}.  At the time of nucleation, $\tau=0$, the vacuum \eqref{falseDplus1} is foliated throughout the $\rho$ direction by surfaces of topology $S^1\times S^{d-1}\times S^{D-d-1}$.  Meanwhile, in the instanton geometry \eqref{bubblenothingKK}, although the asymptotic geometry is the same, by examining the behavior near the bubble \eqref{surfacebubbleAdS} we see that the topology is given by $\mathbb{R}^2 \times S^{d-1}\times S^{D-d-1}$. The change of topology is intimately related to the fact that the instability has been obtained by a double analytical continuation of a non-extremal black hole and it has very interesting consequences when one introduces spinors. Particularly, the loss of the $S^1$ factor between the topology of the vacuum and the bubble geometry represents to decay from a space that is not simply connected, to one that is.  While a simply connected space has a unique choice of spin structure, the vacuum has non-trivial fundamental group and therefore spinors are defined up to a phase $\alpha$
\begin{equation}
 \psi(x,\phi)=\sum_n\,\psi_n(x)\,e^{\frac{i}{\ell_{\text{KK}}} \left(n-\frac{\alpha}{2\pi}\right)\phi}\,.
\end{equation}
The existence of covariantly constant spinors imposes $\alpha=0$ while generally, the unique spin structures on spheres requires that $\alpha=\pi$ (anti-periodic boundary conditions) \cite{Witten:1981gj,Witten:1998zw} in the instanton geometry. This consideration implies that the instanton geometry does not contribute to the path integral of theories described by covariantly constant spinors. Therefore, with suitable boundary conditions on fermions, we can ``cure" the quantum instability of the vacuum \eqref{falseDplus1} and, moreover the ``stable'' boundary conditions correspond exactly to the requirement of the existence of covariantly constant spinors.

As already mentioned, the bubble geometries \eqref{bubblenothingKK} describe (locally) AdS spacetimes in their asymptotics when $\Lambda<0$. The situation including a $S^{D-d-1}$ foliation is particularly interesing since some contributions to the geometry associated with this further foliation are present at the boundary. This implies that, even though the asymptotic geometry is locally $\mathrm{AdS}_{D+1}$, globally it is not. This fact strongly suggests the possibility to extend these backgrounds to theories in which some $p$-form gauge fields are allowed to wrap the spheres realizing the foliations. In this sense one could provide a suggestive interpretation of these non-perturbative decays of $\mathrm{AdS}_{D+1}$ vacua into bubbles of nothing by starting from non-extremal $p$-branes and double-analitically continuing them, in analogy to what is usually done with non-extremal black holes.
Such an analysis, if performed in 10d or 11d supergravities, could provide strong insights regarding the nature of the instabilities of string vacua in relation to de Sitter geometries describing the dynamics of their decays.

\section*{Acknowledgements}

We would like to thank A. Amariti, L. Cassia, S. Klemm, A. Tomasiello for very interesting discussions. NP was partially supported by the Scientific and Technological Research Council of Turkey (T\" ubitak) Grant No.116F137. GD and NP were partially supported by the Principado de Asturias through the grant FC-GRUPIN-IDI/2018/000174. MS is supported by the Swedish Research Council (VR).

\appendix

\section{Hamilton-Jacobi formulation}
\label{HJappendix}

In this appendix we summarize the Hamilton-Jabobi formulation of classical dynamical systems. This formulation of classical mechanics turns out to be very useful to construct solutions in (super-)gravity since it provides a `first-order formulation'' for (non)-extremal gravity backgrounds (see for example \cite{Ceresole:2007wx,Andrianopoli:2009je,LopesCardoso:2007qid,DallAgata:2010ejj,Trigiante:2012eb,Klemm:2016wng,Apruzzi:2016rny,Klemm:2016kxw,Klemm:2017pxv}). In particular the main idea consists in formulating the variational principle of a system by rewriting the action as a sum of squares. The vanishing conditions of each of these squares constitutes a set of first-order differential equations whose solutions are automatically solutions of the equations of motion. More concretely, let's consider a system described by the following action
\begin{equation}
 \begin{split}
  &S(q,\lambda)=\int{d\lambda}\,L(q,\dot q)\,,\\
  &L(q,\dot q)=\frac12 \mathcal{G}_{ij}\dot q^i \dot q^j-V(q)\,,
  \label{lagclasmec}
 \end{split}
\end{equation}
where $\lambda$ is the ``time'' parameter and $(q^i, \dot q^i)$ are the dynamical variables of the system, i.e. the coordinates of the configuration spaces. By Legendre transform one can easly deduce also the the hamiltonian
\begin{equation}
 \begin{split}
  &H(p, q)=\frac12 \mathcal{G}^{ij}p_i p_j+V(q)\,,\\
  &p_i=\frac{\partial L}{\partial \dot q^i}=\mathcal{G}_{ij}\dot q^i\,.
 \end{split}
\end{equation}
The entire method is based on the introduction of a function of the coordinates $F(q)$ called ``superpotential'' or ``Hamilton-Jacobi principal function'' encoding all the informations needed to describe the dynamics of the system. In particular $F(q)$ is defined by the following conditions
\begin{equation}
 \begin{split}
  &H(\partial_q F, q)+\frac{\partial S}{\partial \lambda}=\frac12 \mathcal{G}^{ij}\partial_i F\partial_j F+V-E=0\,,\\
   & p_i=\partial_i F\,,\\
   \label{generalHJ}
 \end{split}
\end{equation}
where $S(q)=F(q)-\lambda E$ with $E$ constant. The first equation is called ``Hamilton-Jacobi equation'' and can be used to rewrite the action as a sum of squares. In particular if we use it to express $V(q)$ in \eqref{lagclasmec} in terms of $F(q)$ and we complete the squares, we get
\begin{equation}
 S=\int{d\lambda\,\left[ \frac12 \,\mathcal{G}_{ij}\,\left(\dot q^i-\mathcal{G}^{ik}\partial_k F\right)\left(\dot q^j-\mathcal{G}^{jl}\partial_l F\right)+\frac{d}{d\lambda}\left(F-E\lambda\right)\right]}\,.
\end{equation}
Up to a total derivative, it is immediate to see that the solutions of the system of first-order equations,
\begin{equation}
 \dot q^i-\mathcal{G}^{ij}\partial_jF\,,
\end{equation}
extremize the action and then are solutions also the equations of motion. We point out that when this method is applied in field theory, the dynamical variables describing a particular configuration of fields have to be considered as the $q$ coordinates of a classical system. Once one is able to find an ``effective lagrangian\footnote{This is not always an immediate procedure. In many cases, especially when gauge fields are included, the 1d effective lagrangian cannot be obtained just by plugging the ansatz for the fields inside the total action.}'' reproducing the equations of motion, the resolution of the Hamilton-Jacobi equations will produce automatically the first-order equations. The crucial point is that the parametrization used to solve the Hamilton-Jacobi equations will determine the explicitly form of the first-order equations. This is coerent with the supersymmetric case in which the SUSY variations of fermions are explicit dependent on the parametrization of the background.

\section{First-order formulation for dS foliations}
\label{bubbleAdS}

In this section we would like to present the explicit derivation\footnote{We point out that the perspective presented in this Appendix may be of interest as a toy model for the research of non-extremal solutions in dimensional supergravity.} of the geometries \eqref{bubblenothingeuclid} via the Hamilton-Jacobi formulation introduced in Appendix \ref{HJappendix} As mentioned in Sec. \ref{discussion}, one can study the decay of the $(D+1)$-dimensional vacua \eqref{falseDplus1} by considering the dynamics of a scalar field coupled to a $D$-dimensional singular background. In particular, a good lower-dimensional description of the bubble geometries \eqref{bubblenothingKK} can be obtained by setting to zero the KK vector and just by considering a scalar field $\Phi$ in $D$ dimensions.
We will start by connecting a class of $D$-dimensional backgrounds of the type,
\begin{equation}
\begin{split}
 &ds_D^2=dr^2+e^{2U(r)}\left(L_{{\scriptsize \mrm{dS}}_d}^2ds^2_{{\scriptsize \mrm{dS}}_d} +L_{S^{D-d-1}}^2 ds^2_{S^{D-d-1}} \right)\,,\\
 &\Phi=\Phi(r)\,,
 \label{dSAdS}
 \end{split}
\end{equation}
to the $(D+1)$-dimensional KK vacuum by reducing on a circle via the KK ansatz,
\begin{equation}
 \begin{split}
  &ds^2_{D+1}=e^{2 \alpha_D \Phi}ds^2_D+e^{2 \beta_D \Phi}d\phi^2\,,\\
  &\alpha_D= \frac{\kappa_D }{\sqrt{2(D-1)(D-2)}}\qquad \text{and} \qquad \beta=-(D-2)\alpha_D\,,
  \label{KKansatzAdS}
 \end{split}
\end{equation}
where the coordinate $\phi$ is periodic with $\phi \sim \phi +2 \pi \,\ell_{\text{KK}}$ . 
The corresponding $D$-dimensional action has the form
\begin{equation}
 S_D=\frac{1}{2\kappa_D^2}\int{d^D x}\,\sqrt{-g_D}\,\left (R_D-\frac{\kappa_D^2}{2}\, \left(\partial \Phi \right)^2-2 V(\Phi)\right)\,,
 \label{DdimGravityDilatonAdS}
\end{equation}
where $\kappa_{D+1}^2=2\pi\,\ell_{\text{KK}}\,\kappa_D^2$ and
\begin{equation}
 V(\Phi)=-\frac{D(D-1)}{2\,\ell^2}\,e^{2\alpha_D \Phi}\,.\label{potential}
\end{equation}
Here, and in the following we specialize to the case $\Lambda <0$, however the $\Lambda=0$ case can always be recovered by taking $l\to \infty$.

Let's find an explicit solution for the $D$-dimensional backgrounds \eqref{dSAdS} following the procedure outlined in Appendix \ref{HJappendix}. 
As we mentioned in \secref{bubbleKK}, the consistency of the equations of motion of  \eqref{DdimGravityDilatonAdS} requires that we need to impose the following constraint on the radii of $\mrm{dS}_d$ and of $S^{D-d-1}$,
\begin{equation}
L_{S^{D-d-1}}=L_{{\rm dS}_d}\sqrt{\frac{D-d-2}{d-1}}\,.
 \label{radiiconstraint1}
\end{equation}
One can  show that the equations of motion of \eqref{DdimGravityDilatonAdS} can be reproduced by the following effective Lagrangian
\begin{equation}
 L_{eff}=e^{(D-1)U}\left((D-1)(D-2)(U^\prime)^2-\frac{\kappa_D^2}{2}(\Phi^\prime)^2+\frac{(d-1)(D-1)}{L_{{\rm dS}_d}^2}e^{-2U}-2V(\Phi)\right)\,.
\end{equation}
 Introducing the conjugate momenta $\pi_U=\partial_{U^\prime}L_{eff}$ and $\pi_\Phi=\partial_{\Phi^\prime}L_{eff}$, one immediately obtains the effective hamiltonian of the system,
\begin{equation}
 H_{eff}=e^{-(D-1)U}\left( \frac{\pi_U^2}{4(D-1)(D-2)}-\frac{\pi_\Phi^2}{2\kappa_D^2}\right)-\frac{(d-1)(D-1)}{L_{{\rm dS}_d}^2}e^{(D-3)U}+2 V(\Phi)e^{(D-1)U}\,.
 \label{heffdSKK}
\end{equation}
Comparing $H_{eff}$ with the equations of motion, we see that the Hamiltonian constraint has the form $H_{eff}=0$.
By introducing a prepotential $F=F(U,\Phi)$ such that $\pi_U=\partial_{U}F$ and $\pi_\Phi=\partial_{\Phi}F$, we obtain the Hamilton-Jacobi equation by specifying the general expression \eqref{generalHJ}
\begin{equation}
  \frac{(\partial_U F)^2}{4(D-1)(D-2)}-\frac{(\partial_\Phi F)^2}{2\kappa_D^2}-\frac{(d-1)(D-1)}{L_{{\rm dS}_d}^2}e^{2(D-2)U}-\frac{D(D-1)}{\ell^2}e^{2(D-1)U+2\alpha_D \Phi}=0\,.\label{HJeqdSAdS}
\end{equation}
A solution of \eqref{HJeqdSAdS} can be obtained by taking inspiration from \cite{Klemm:2017pxv} where the Hamilton-Jacobi equation is solved for general non-extremal $\mrm{AdS}$ black holes in the Einstein-Maxwell theory. In particular one can verify that 
\begin{equation}
 F(U, \Phi)=c\,e^{(D-2)U}\left(e^{-\alpha_D (D-2)\Phi}+\nu\, e^{\alpha_D (D-2)\Phi} \right)+\frac{(D-1)^2}{c\,l^2}e^{D(U+\alpha_D\Phi)}\,,\label{AdSsuperpot}
\end{equation}
where $c$ is a real integration constant and 
\begin{equation}
 \nu=\frac{(d-1)(D-1)^2}{c^2L_{{\rm dS}_d}^2(D-2)}\,.\label{nu}
\end{equation}
We point out that a generic choice of $c$ gives rise to one-parameter family of solutions from the $D$-dimensional point of view. However only a particular value of this constant will reproduce the $(D+1)$-dimensional smooth bubble geometry written in \eqref{bubblenothingKK}.

The first-order equations can be easly obtained by comparing the definition of the conjugate momenta, $\pi_U=\partial_{U^\prime}L_{eff}$ and $\pi_\Phi=\partial_{\Phi^\prime}L_{eff}$, with the expression $\pi_U=\partial_{U}F$ and $\pi_\Phi=\partial_{\Phi}F$. In this way one obtains
\begin{equation}
\begin{split}
& U^\prime=\frac{e^{-(D-1)U}}{2(D-1)(D-2)}\,\partial_U F(U,\Phi)\,,\\
 &\Phi^\prime=-\frac{e^{-(D-1)U}}{\kappa_D^2}\,\partial_\Phi F(U,\Phi)\,.
 \end{split}\label{floweqdSAdS}
\end{equation}
Finally one can show that \eqref{floweqdSAdS} imply the all the equations of motion.
The equations \eqref{floweqdSAdS} can be intregrated out by introducing the new dynamical variable
\begin{equation}
 \rho=e^{U+\alpha_D \Phi}\,.\label{AdSrhodef}
\end{equation}
In this way, by taking the ratio of the two equations \eqref{floweqdSAdS}, we get
\begin{equation}
 \frac{d\Phi}{d\rho}=-\frac{1}{2\alpha_D\rho}+e^{2(D-2)\alpha_D \Phi}\left(\frac{\nu}{2\alpha_D\rho}+\frac{D(D-1)^2}{2c^2\,\ell^2\,\alpha_D(D-2)}\,\rho   \right)\,,
\end{equation}
that it is solved by
\begin{equation}
 e^{\Phi}=\left (\nu+ \frac{(D-1)^2}{c^2\,\ell^2}\,\rho^2+\frac{C_0}{\rho^{D-2}}\right)^{-\frac{1}{2(D-2)\alpha_D}}\,,
\end{equation}
where $C_0$ is a positive integration constant. From \eqref{AdSrhodef} we can work out the expression for $U$,
\begin{equation}
e^U=\rho \, \left (\nu+ \frac{(D-1)^2}{c^2\,\ell^2}\,\rho^2+\frac{C_0}{\rho^{D-2}}\right)^{-\frac{1}{2(D-2)}}\,,
\end{equation}
and, finally, by using \eqref{KKansatzAdS} to uplift to $D+1$ dimensions, we obtain the following background
 \begin{equation}
 \begin{split}
 ds^2_{D+1}&=\left(\nu+\frac{(D-1)^2}{c^2\,\ell^2}\,\rho^2+\frac{C_0}{\rho^{D-2}}\right)d\phi^2+\left(\frac{(D-1)^2}{c^2}\right)\frac{d\rho^2}{\nu+\frac{(D-1)^2}{c^2\,\ell^2}\,\rho^2+\frac{C_0}{\rho^{D-2}}}\\
 &+\rho^2L_{{\rm dS}_d}^2ds^2_{{\scriptsize \mrm{dS}}_d\times S^{D-d-1}}\,,
 \label{bubblenothingAdSsol}
 \end{split}
\end{equation}
where $ds^2_{{\scriptsize \mrm{dS}}_d\times S^{D-d-1}}=ds_{{\rm dS}_d}^2 +\left(\frac{D-d-2}{d-1} \right) ds^2_{S^{D-d-1}}$. If we require that the metric \eqref{bubblenothingAdSsol} has to reproduce the $(D+1)$-dimensional Minkowski vacuum when $C_0=0$ and $\ell\rightarrow +\infty$, we obtain that
\begin{equation}
  \nu=1 \Longleftrightarrow c=(D-1)\sqrt{\frac{d-1}{D-2}}L_{{\rm dS}_d}^{-1}\,.
\end{equation}
From this condition we can finally recast the metric into the following form
 \begin{equation}
 \begin{split}
 ds^2_{D+1}&=\left(1+\left(\frac{D-2}{d-1}\right)\,\frac{\rho^2}{l^2}-\left(\frac{R}{\rho}\right)^{D-2}\right)d\phi^2+\left(\frac{D-2}{d-1}\right)\frac{d\rho^2}{1+\left(\frac{D-2}{d-1}\right)\,\frac{\rho^2}{l^2}-\left(\frac{R}{\rho}\right)^{D-2}}\\
 &+\rho^2ds^2_{{\scriptsize \mrm{dS}}_d\times S^{D-d-1}}\,,
 \label{bubblenothingAdS}
 \end{split}
\end{equation}
where we absorbed the dependence on $C_0$ and $L_{{\rm dS}_d}$ by introducing the positive parameter $R$.
It is manifest that the solution \eqref{bubblenothingAdS} reproduces \eqref{bubblenothingKK} in for $\Lambda <0$.  Therefore the vacuum decay via a bubble of nothing can be captured using a tunneling scalar  field in one less dimension. 



 \bibliographystyle{utphys}
  \bibliography{references}
\end{document}